# Encouraging Knowledge Sharing using Web 2.0 Technologies in Higher Education: A Survey

Shuaibu Hassan Usman[1] and Ishaq Oyebisi Oyefolahan[2]

[1]Department of Management and Information Technology, Abubakar Tafawa Balewa University, Bauchi, Nigeria
[2]Department of Information System, International Islamic University Malaysia, Kuala Lumpur, Malaysia

## ABSTRACT

*As the technology continuous to advance, new technologies have emerged with the capability to revolutionize knowledge sharing practices. Web 2.0 exemplifies such new technologies, which provides dynamic way of interactions of people and businesses. In learning environment, Web 2.0 technologies support and enhance teaching and learning of students. Therefore, the main aim of this study focuses on the determining the ways to encourage knowledge sharing through web 2.0 technologies from students' point of views. A total of 287 students responded to the online questionnaire in International Islamic University Malaysia (IIUM). Descriptive statistics was used in data analysis. The results show that students used web 2.0 technologies in learning and sharing knowledge among them. In addition, the study found eight items on ways to encourage and enhance knowledge sharing among students in the University. These items include Create Awareness, Provide facilities, Internet Accessibility, Ease of use, Encourage Teamwork, Materials Availability, Improved and Response, and Motivation.*

## 1. INTRODUCTION

Web 2.0 technologies provide dynamic ways of interactions of people and businesses. The web 2.0 sometimes refers to 'read and write', 'create and share', 'like and comment', 'customer content creator' and many more names. All these names are possible due to the crucial transformational role this web 2.0 plays in  changing the way people interact, communicate, collaborate and share information among friends, families, co-workers, and even strangers. The transformation has made the sharing of Knowledge possible through sharing of information and experiences of people across geographical disperse and of different social and cultural background.

In today's education, the emphasis of learning has shifted to students learning centric. For instance, Don Tapscott [1] maintains that a new model of education has started to emerge, which is student-focused. The model of pedagogy that is student-focused centres more on collaboration (i.e. between lecturers and students) and multi-way, which is customized [1]. For that reason, the web 2.0 technologies and beyond have become the backbones to facilitate learning among students in 21st century educational environment. Looking at the web 2.0 technologies in the context of knowledge sharing among students, there is a need to determine the extent of its use and ways of improving it in higher education. Therefore, this study focuses on two objectives as under:

1. To find out the purposes for which the students used the web 2.0 technologies in knowledge sharing.
2. To determine the ways to improve knowledge sharing with web 2.0 technologies among student in universities.

## 2. THE WEB 2.0 TECHNOLOGIES AND LEARNING ENVIRONMENT

Web 2.0 signifies the development in the use of World Wide Web applications and designs. The web 2.0 combines the concepts, technologies, and trends that enable users to shares, connects, communicates, collaborates, and creates information on the web. The term 'web 2.0' is credited to Tim O'Reilly following a conference on the next-generation Web concepts and issues held by O'Reilly Media and Media Live International in 2004 [2]. The web 2.0 is a new version of the World Wide Web that shows the changes in the ways end-users and software developers utilize the Web without referring to an upgrade or update to any technical specifications [2].

The major breakthrough of web 2.0 technologies is that they are inexpensive and available to individuals who have Internet access, and allow them to be producers as well as be among the worldwide learning community that connects, communicates, collaborates, and shares information [3]. The main advantage of Web 2.0 over the previous websites is that it does not require technical expertise such as web design or publishing skills to contribute and play a part, as such it has made easy for people to create, publish, collaborate, and communicate their work or research with others across the globe [4]. The web 2.0 technologies comprise of wikis, blogs, social networking, podcasting, content hosting services, bookmarking sites etc., which support and promote the concept of shared knowledge. Therefore, these types of technologies are influencing the 21st century learning environments. For instance, universities can use the web 2.0 applications to communicate with students, staff and the wider academic community. In addition, the technology has provided a platform for research collaboration and communication among students as well as staff [4]. According to Hew and Cheung [5], history has shown the concern that educators have in finding a suitable technology to be used for enhancement of student learning and fortification of education. They conclude that web 2.0 is among the recent technologies, which has captured the attention of many educators around the world. They posit that web 2.0 is read-write web, which allows two-way communication between the site and users. Furthermore, web 2.0 tools allow individuals to collaborate and contribute with one another to the authorship of content, customize web sites for their use, and instantly publish their opinion. With web 2.0 technologies, individuals can become producers and contributors to web sites instead of merely being consumers of others' work. As a result, there are several types of web 2.0 technologies available for institutions to use with their staff and students [5].

## 3. RELATED STUDIES

A study of John thompson (2007) indicated that Web 2.0 has positively influenced education by providing an instant two-way platform of web contents dissemination and creation. In the past, students interact with the web resources merely as consumers and receivers without any creativity or contribution. However, the revolutionary activities of web 2.0 have changed this game for students and users as now students and users actively revolve around web contents such as commenting, posting, and uploading the contents [6].

Similarly, a study found that providing facilities encourage and help students to share knowledge using electronic storage and other platforms in a simpler access as well as encouraging economic reuse of knowledge [7]. The author continues saying that "*IT systems can provide codification, personalization; electronic repositories for information and can help people locate each other to communicate directly*" [7]. In a nutshell, availabilities of information technology (IT) tools and applications provide a platform to enhance and facilitate knowledge sharing and knowledge management [8] [7].

Equally, success in knowledge sharing requires some conditions to be fulfilled that were listed in the study of Frost [7] who cited Bukowitz and Williams (1999) as under:

1. **Articulation:** The ability of the user to define what he needs.

2. **Awareness:** Awareness of the knowledge available. The provider is encouraged to make use of directories, maps, corporate yellow pages, etc.

3. **Access:** Access to the knowledge.

4. **Guidance:** Knowledge managers are often considered key in the build-up of a knowledge sharing system. They must help define the areas of expertise of the members of the firm, guide their contributions, assist users, and be responsible for the language used in publications and other communication material. This is so as to avoid an information/knowledge overload.

5. **Completeness**: Access to both centrally managed and self-published knowledge. The former is often more scrutinized but takes longer to publish and is not as hands-on (and potentially relevant). Self-published information on the other hand runs the risk of not being as reliable [7].

In essence, Web 2.0 technologies have dramatically changed the way people interact with information and data resources on the internet because they allow conversation on the contents and information publication. As explains by S. Hargadon [9], the web has allowed information publication to attract comments and suggestions from various users, which help to overcome the problem of Information overload. According to the author, "*When, however, we see the ever increasing amount of content as 'conversations' that are taking place, it becomes an educational imperative to teach ourselves and students to be productive participants in those conversations… the answer to information overload is to create (and to teach the creation of) more information*" [9].

## 4. METHODOLOGY

This study used a semi-structured questionnaire in data collection from students of International Islamic University Malaysia (IIUM). The questionnaire was administered online through Google Document in which 287 students responded. The study is a survey in nature that comprises of elements of both quantitative and qualitative data. In the part of data analysis, a simple descriptive statistics was used to analyse the data.

## 5. IMPLICATION OF THE STUDY

The study has explored the ways to enhance knowledge sharing within the university. Therefore, the implication of this study contributes toward policy making regarding the existing technologies provided by the university. The university has web 2.0 technologies such as learning management systems, Claroline in KICT, and other social media groups with solid aim of facilitating knowledge sharing among learners. With this study, the university has an insight on some issues that contribute to knowledge sharing using web 2.0 tools among the students. The university can use this study to formulate strategies, which will carry all students along in order to meet the objectives of the institution.

## 6. RESULTS

The descriptive statistics for quantitative data demonstrates the results of specific usage behaviour of the students with the web 2.0 technologies. The statement '**Daily with my colleagues**' has 51 (17.8%) responses; the second statement '**Daily to learn or share**' has attracted 60(20.9%) responses; the third statement '**Weekly in class**' has 11(3.8%); the fourth statement '**I use it when I have assignment, class project, term paper, presentation etc**.' has 161 (56.1%); and the last statement '**Other**' has 4(1.4%). Overall, the statement '**I use it when I have assignment, class project, term paper, presentation etc'** has the highest respondents, which indicates the students use the tools mostly for academic purpose. Interestingly, the statement '**Daily to learn or share'** has a substantial percentage of the respondents who use the tools for both learning and sharing of ideas. Table 1 depicts these results.

Table 1: Description Statistic of Usage Behaviour with web 2.0 technologies

| Specific Usage Behaviour of the Students with the Tools | Frequency | Percentage (%) |
|---|---|---|
| Daily with my colleagues | 51 | 17.8 |
| Daily to learn or share | 60 | 20.9 |
| Weekly in class | 11 | 3.8 |
| I use it when I have assignment, class project, term paper, presentation etc. | 161 | 56.1 |
| Other | 4 | 1.4 |

However, the results of qualitative data show eight distinct groups items based on keyword. The students were asked a question 'What are the ways to improve knowledge sharing using web 2.0 technologies in IIUM? The responses show a wide range of suggestions that are depicted in the table 2.

1. **Create Awareness:** Many students suggested that the university should formulate ways to create awareness on how these tools and their features can be utilized. They believe that there are gaps existing among students, lecturers, and faculties or departments on issues of using ICT tools in most cases.

2. **Provide facilities**: the students suggest that the university needs to incorporate some facilities to the existing tools as they believe there are more facilities to be provided to support and enhance students' learning and sharing.

3. **Internet Accessibilities:** Almost all the students have made comments regarding to the speed and coverage of the internet connections in the university. Although, the university has made a tremendous effort in providing high speed internet connection with two devices from both students and staff to be connected, still some areas have limited internet coverage especially Muhallahs (hostels). The limited internet coverage has drawn the attention of the students to make suggestion.

4. **Ease of use:** Many of the students are of the opinion to improve the user interface to enable easy navigation and use by both novice and expert students. The design of the existing tools tailored more on those that have the skills and expertise on IT based on the opinion of some students.

5. **Encourage Teamwork**: Suggestion on ways to improve knowledge sharing in IIUM from the respondents show that teamwork plays important role. Teamwork is essential among students as well as between students and lecturers because it unites various diverse skills from many students.

6. **Materials Availabilities**: Many students have traced the need for adequate provision of learning materials such videos and class contents online via these web 2.0 tools for off the class revision and continuous replay. By and large, students value the convenience, choice, and flexibility of the online material. This is because they can significantly learn and share with each other easily. Therefore, making the materials available can support student learning and increase student achievement and success.

7. **Improved Responses:** some of the students suggest that the tools in IIUM should be made powerful to accommodate the features of communication among students, and between students and lecturers.

8. **Motivation**: The responses suggested that motivation help students to develop interest in an activity. This can make them to give more attention and show signs of more enthusiasm to perform the task as it motivates them to gain confidence and self-esteem.

Table 2: The Ways to Improve Knowledge Sharing in IIUM

| Questionnaire Item | Responses (decoding) | Theme |
|---|---|---|
| Please suggest ways to improve knowledge sharing in IIUM | a. Organize a workshop for students to learn how to use the features of E-learning in IIUM | Create awareness |
| | b. Provide online group discussion to have a good network connection and frequent workshop | |
| | c. Create awareness to students through portal | |
| | d. To conduct a briefing session regarding IIUM knowledge sharing tools. | |
| | e. Awareness and support to students, an organized public online forum can benefit us. | |
| | f. Create awareness on knowledge sharing in IIUM especially to students without ICT background like law students | |
| | g. Publicize the tools to student and attach with user manual, each Faculty should have E-learning like KICT possibly | |
| | h. … Organizing public talk and speaking on the importance of sharing ideas. | |
| | i. Organize workshop, conferences and talks to students and lecturers to enable them familiar with the KS tools and every Kulliyah need to have a benchmark for KS | |

Table 2                    continued

| Questionnaire Item | Responses (decoding) | Theme |
|---|---|---|
| Please suggest ways to improve knowledge sharing in IIUM | 1. Proper tool and training | Provide facility |
| | 2. Use edmodo- it is relevant, attractive, easy to use, and easy to get responses | |
| | 3. Instance message and chat apps should be added to LMS to enable students to interact with their lecturers directly | |
| | 4. Provide audio & video conferencing | |
| | 5. The IIUM should make computer labs for every kulliyah available 24hour a day, 7days a week to enable those students without computers to do their assignments and share knowledge | |
| | 6. Upgrading the LMS to moodle so that everyone in the class can chat and discuss each other | |
| | 7. Create a forum for collaboration and communication between staff and students | |
| | 8. Instant messaging facilities that will allow students to reach their lecturers directly | |
| | 9. Provide online edition tools to enable students edit files together online | |
| | 10. IIUM should try to create the newest facilities which will make its student up to date | |
| | 11. Free access to computers in every class or studio for those who need to do research/use the computer | |
| | 12. Provide more interactive and lively feature. | |
| | 13. Enable calling facilities on IT tools between students and lecturers | |
| | 14. Some students possess some skills like software code development and simulation that are beneficial to other students as such there is a need for a online forum to encourage sharing of these technical skills | |
| | 15. IIUM need to add features like IM, skype, and so on | |
| | 16. Provide Apps for student in IIUM | |
| | 17. Establish online forum to make the discussion easier among students provide a single platform for students to easily communicate with each other, and report should be generated to enable deans and HOD to have access to it | |

Table 2                    continued

| Questionnaire Item | Responses (decoding) | Theme |
|---|---|---|
| Please suggest ways to improve knowledge sharing in IIUM | a. Network performance need to be efficient, so provide high speed internet | Internet Accessibility |
| | b. Uninterrupted and speedy Internet connectivity | |
| | c. I hope the highest authority in IIUM will take an aggressive action about the problem related to Wi-Fi because students are having problem of finishing their works/assignments as the internet connection always makes troubles, which affect learning and sharing of knowledge in IIUM | |
| | d. Improve internet connection | |
| | e. Upgrade the speed and improve Wi-Fi | |
| | f. Improve Wi-Fi especially in Muhallah | |
| | g. Reliable access to internet | |
| | h. … smooth running internet at all allocations in the campus is an important factor that will help students to work anytime to finish their assignments before the deadline for submission as lecturers feel unconcern on challenges for failure to meet schedule | |
| | | |
| | 1. More features to existing tools to support learners | Ease of Use |
| | 2. Make the User Interface more user friendly and have good maintenance of its availability | |
| | 3. Make it more user-friendly and mobile devices Compatible | |
| | 4. Use interactive media to make class more enjoyable | |
| | 5. The User Interface should be improved | |
| | 6. Improve the design of the web and give more option for student to upload things there. Sometimes lecturers limit the functionality for students to do that. Maximize the student control in the learning environment | |
| | 7. Improve the User Interface and use multimedia content | |
| | 8. Make it user-friendly to students | |
| | 9. The e-learning should be improved with better user interface so that students will be able to use it | |
| | 10. Make it more easy and friendly to share ideas | |
| | 11. Regular make the e-learning more interactive more user-friendly | |

Table 2                     continued

| Questionnaire Item | Responses (decoding) | Theme |
|---|---|---|
| Please suggest ways to improve knowledge sharing in IIUM | a.  Encourage group assignment | Encourage Teamwork |
| | b.  Forcing students using it by requiring teamwork assignments through it or extra marks for who participate there | |
| | c.  Do more on group assignment | |
| | d.  Teamwork should be encourage | |
| | e.  Lecturers should encourage group work | |
| | 1)  Access to LMS/e-learning. Besides, announcement and document should be made accessible by mobile apps | Materials Availability |
| | 2)  The university should encourage recording of lectures in video format and make them available for students online develop an apps with direct | |
| | 3)  Web based questions and answers should be incorporated to the Knowledge sharing tools in IIUM | |
| | 4)  Make previous research work available to student | |
| | 5)  Record the class's lectures and tutorial for later use by students. they can able to play it back when they need it | |
| | 6)  Lecturers should use audio/video to encourage sharing of ideas | |
| | 7)  widen the sharing of Islamic literature | |
| | 8)  By sharing the video lectures of the classes | |
| | 9)  Create online archive of articles, lectures, and notes | |
| | 10) Lecturers need to have their own blog regarding to their subject teaching. This will enable students to comments and ask questions and have responses from the lecturer. This will eliminate the physical consultation which has limit time and place | Improved Responses |
| | i.  The tools should be designed for real-time knowledge sharing between staff and students | |
| | ii.  Close the gap between students and lecturers, and expose students to online sharing of ideas | |
| | I.  More encouragement through IIUM media | Motivation |
| | II.  Lecturers should use audio/video to encourage sharing of ideas | |
| | II.  Encourage students to use blog to share ideas | |
| | V.  Encourage students to focus on academic conversations | |
| | V.  Motivate students to share their ideas | |
| | I.  Lecturer should motivate students to use the tools | |

# 7. DISCUSSIONS

The descriptive statistics results confirmed the use of web 2.0 tools among students in IIUM Gombak campus to accomplish their academic activities. This finding is comparable to that of Muñoz & Towner [10] where their study revealed that students are committed to the use of Web 2.0 technologies (i.e. blogs, twitter, podcasts, wikis, social network sites, virtual worlds, video sharing and photo sharing) [10]. In this study, the students indicate that they are performing various activities that involve both formal and informal learning. From the results, it can be seen that the students in the university use these tools for academic purposes more than other social aspects as the statements 'I use it when I have assignment, class project, term paper, presentation etc' and 'Daily to learn or share' have attracted high responses from the students. Similarly, this research is accordance with a study, which found three effective educational technologies and useful of blogs among students that include reading blogs of others, receiving comments, and previewing tasks of others and reading feedback received in relation to these [11].

Lastly, a qualitative results show the eight items on ways to encourage and enhance knowledge sharing among students in the University. These items include Create Awareness, Provide facilities, Internet Accessibility, Ease of use, Encourage Teamwork, Materials Availability, Improved Communication and Response, and Motivation. This implies that the university needs to do more to encourage knowledge sharing among students.

# 8. CONCLUSIONS

To enhance the quality of knowledge sharing using web 2.0 among students in IIUM, the university needs to consider the suggestions provided by the students. The participants have suggested ways to encourage knowledge sharing across the university. These ways include communicate features, collaboration features, creating awareness and training, providing learning facilities and materials, fast internet connectivity, and motivation. It is essential for the university to recognize the various ways leading to knowledge sharing among students using web 2.0 tools. Therefore, the future work of this research will be on the web 2.0 enabled knowledge sharing activities among student in the university using case study approach. This approach will provide an in depth study of knowledge sharing across the entire faculties of the university to find out the factors that influence students to share their knowledge with their colleagues and others.